\documentclass[12pt]{iopart}
\usepackage{iopams}  
\usepackage{graphicx}

\begin{document}
\title{Low temperature optical characterization of near infrared single photon emitters in nanodiamonds}
\author{P~Siyushev$^{1}$, V~Jacques$^{1,{\ast}}$, I~Aharonovich$^{2}$, F~Kaiser$^{1}$, T~M$\ddot{\rm u}$ller$^{3}$, L~Lombez$^{3}$, M~Atat$\ddot{\rm u}$re$^{3}$, S.~Castelletto$^{2}$, S~Prawer$^{2}$, F~Jelezko$^{1}$ and J~Wrachtrup$^{1}$}

\vspace{1cm}

\address{$^{1}$
3.Physikalisches Institut, Universit$\ddot{\rm a}$t Stuttgart, D-70550 Stuttgart, Germany}
\address{$^{2}$ School of Physics, University of Melbourne, Victoria 3010, Australia}
\address{$^{3}$ Cavendish Laboratory, University of Cambridge, Cambridge CB3 0HE, United Kingdom}

\address{$^{\ast}$ Corresponding author v.jacques@physik.uni-stuttgart.de}

\begin{abstract}
In this paper, we study the optical properties of single defects emitting in the near infrared in nanodiamonds at liquid helium temperature. The nanodiamonds are synthesized using a microwave chemical vapor deposition method followed by nickel implantation and annealing. We show that single defects exhibit several striking features at cryogenic temperature: the photoluminescence is strongly concentrated into a sharp zero-phonon line in the near infrared, the radiative lifetime is in the nanosecond range and the emission is perfectly linearly polarized. The spectral stability of the defects is then investigated. An optical resonance linewidth of $4$ GHz is measured using resonant excitation on the zero-phonon line. Although Fourier-transform limited emission is not achieved, our results show that it might be possible to use consecutive photons emitted in the near infrared by single defects in diamond nanocrystals to perform two photon interference experiments, which are at the heart of linear quantum computing protocols.
\end{abstract}


\submitto{\NJP}

\maketitle

 \indent Over the last decade, optically-active defects in diamond have been demonstrated as efficient and robust solid-state single photons sources owing to perfect photostability at room temperature~\cite{Kurtsiefer_PRL2000,Beveratos_OptLett2000}. Among many studied color centers in diamond including silicon-vacancy (SiV)~\cite{SiV} and nickel-related NE8 defects~\cite{Gaebel_NJP2004,Wu_OptExp2006}, the nitrogen-vacancy (NV) color center has attracted a lot of interest because its spin state can be coherently manipulated with high fidelity owing to extremely long coherence time~\cite{Gupi_NatMat2009}. Moreover, Fourier-transform limited emission from the zero-phonon line (ZPL) of single NV defect in diamond has recently been reported using resonant excitation at cryogenic temperatures~\cite{Tamarat_PRL2006,Batalov_PRL2008}. Such results make the NV color center a competitive candidate for solid-state quantum information processing (QIP). However, NV defects exhibit a broad spectral emission associated with a Debye-Waller factor on the order of $0.05$, even at low temperature~\cite{Jelezko_APL2002}. Emission of single photons in the ZPL is then extremely weak, typically on the order of a few thousands of photons per second. Such counting rates might be insufficient for the realization of advanced QIP protocols based on coupling between spin states and optical transitions within reasonable data acquisition times. Consequently, it is highly desirable to investigate new defects in diamond which would combine the unprecedented properties of spins in diamond~\cite{Gupi_NatMat2009} and a bright emission in the zero-phonon line. \\
\indent Recent development of chemical vapor deposition (CVD) technique accompanied with alternative fabrication methods of optical centers in diamond revealed a completely new family of defects with a ZPL centered in the near infrared (NIR)~\cite{Igor_APL2008,Igor_PRB2009}. In this work, we study at liquid helium temperature the optical properties of such defects in CVD-grown nanodiamonds. First, we show that they exhibit several striking features: the photoluminescence (PL) is highly concentrated into a sharp zero-phonon line, corresponding to a Debye-Waller factor larger than $0.9$, the radiative lifetime is $2$ ns and the emission is linearly polarized. As most of the applications of single-photon source in the field of quantum information processing requires the emission of Fourier-transform limited single photons~\cite{RevueOrrit,KLM,OBrien_Science2007}, we then investigate the spectral stability of the defects using resonant excitation on the ZPL. Low amplitude spectral jumps are evidenced, leading to an optical linewidth in the GHz range. Although Fourier-transform limited emission is not demonstrated, the spectral diffusion time is measured in the millisecond range, which might be long enough to use consecutive emitted photons to perform two-photon interference experiments.\\

\indent The investigated diamond nanocrystals were grown using a microwave assisted CVD technique on a sapphire substrate. Following a procedure described in Ref.~\cite{Stacey_DRM2009}, the substrate was first seeded with a nanodiamond powder ($4$-$6$ nm size, Nanoamor Inc.) and then loaded into a $900$ W microwave plasma CVD reactor. The chamber gas mixture consisted of $2\%$ CH$_{4}$ in H$_{2}$ and the chamber pressure was maintained at $150$ Torr during the growth
process, with a substrate temperature of $800 \ ^{\circ}$C. After the growth, nickel implantation was performed for introduction of vacancies using a focused ion beam (FIB) loaded with a Ni/Er alloy liquid metal ion source~\cite{Igor_PRB2009}. After implantation, the sample was finally annealed at $1000 \ ^{\circ}$C during one hour in a $95\%$Ar-$5\%$H$_{2}$ ambient. Such a procedure leads to the formation of nanodiamonds with a size of about $200$-$600$ nm, hosting defects with a ZPL emitting in the NIR at a wavelength mainly around $770$ nm~\cite{Igor_PRB2009}, but within a range of $770\pm15$ nm. We note that the defects studied in this paper are not necessarily related to nickel. Indeed, recent results suggest that an incorporation of Cr atoms during CVD growth of diamonds on a sapphire substrate lead to a formation of single photon emitters with a ZPL at $750$-$760$ nm~\cite{Igor_NanoLett2009}. The atomic structure of the studied defects will not be addressed in this study.\\
\begin{figure}[b]
\centerline{\resizebox{0.75\columnwidth}{!}{
\includegraphics{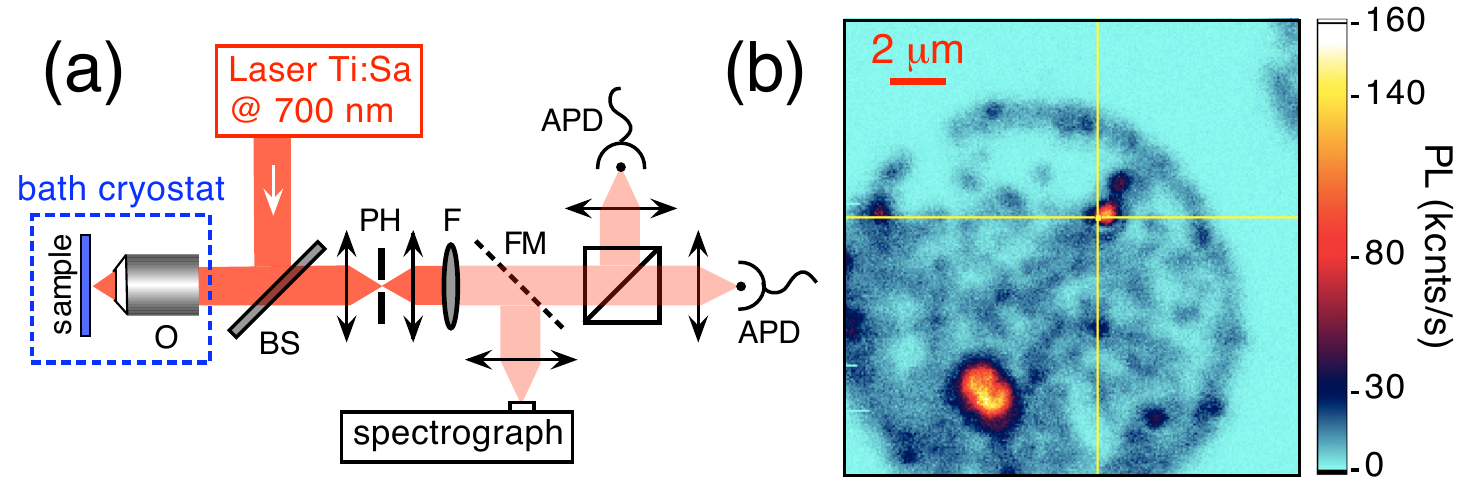}}}
\caption{(a)-Experimental setup. BS: quartz plate with $5\%$ reflectance; O: microscope objective with a numerical aperture of $0.85$ immersed in the helium bath; PH: $100 \ \mu$m  diameter pinhole; F: combination of a $750$ nm long-pass filter and a $785$ nm short-pass filter; FM: flip mirror directing the PL either to an imaging spectrometer (Acton research) equipped with a back-illuminated cooled CCD matrix, or to a Hanbury-Brown and Twiss interferometer consisting of two silicon avalanche photodiodes (APD) placed on the output ports of a $50/50$ beamsplitter. (b)-Typical raster scan of the sample recorded with an excitation power of $300 \ \mu$W. The PL intensity is encoded in color scale and the yellow cursor indicates a single defect emitting in the NIR, imaged with a signal to background ratio on the order of $10$:$1$.}
\label{Fig1}
\end{figure}
\indent Single defects were imaged using conventional confocal microscopy at low temperature ($T\approx 2$ K), as depicted in figure~\ref{Fig1}(a). The optical excitation was carried out using a titanium:sapphire laser (Coherent Mira) operating at the wavelength $\lambda=700$~nm, either in a continuous mode or in a femtosecond pulsed regime at a repetition rate of $76$ MHz. The laser beam was focused on the sample with a microscope objective (NA=$0.85$) immersed in the helium bath cryostat. PL was collected by the same objective and spectrally filtered from the remaining pump light using a $750$ nm long-pass filter combined with a $785$ nm short-pass filter. Following standard confocal detection scheme, the collected light was then focused onto a $100 \ \mu$m diameter pinhole and directed either to a spectrometer or to a Hanbury-Brown and Twiss (HBT) interferometer used for photon correlation measurements. Lateral ($x$-$y$) raster scan of the sample was realized using a scanning galvanometer mirror system combined with telecentric lenses, while $z$-scan was provided by a high resolution piezoelectric stepper (Attocube) on which the substrate  with the CVD-grown nanodiamonds was glued. A typical PL raster scan of the sample is depicted in figure~\ref{Fig1}(b), showing isolated and bright photoluminescent spots which correspond to defects emitting in the NIR in CVD-grown diamond nanocrystals.\\
\indent The PL spectra of a single defect recorded at $2$ K and $300$ K are shown in figure~\ref{Fig2}(a). The emission is highly concentrated into a sharp zero-phonon line (ZPL) at a wavelength around $770$ nm. At room temperature, a tiny phonon side band can be observed around $780$ nm, which is not visible anymore at low temperature. Consequently, the Debye-Waller factor, {\it i.e.} the ratio between the intensity of the ZPL and the total intensity of the emission spectrum, is extremely high ($>0.9$) at liquid helium temperature. For comparison, the PL associated with NV defects in diamond at low temperature exhibits a Debye-Waller factor on the order of $0.05$~\cite{Jelezko_APL2002}. \\
\indent The statistic of the emitted photons was then investigated for the same emitter by recording the histogram of time delays between two consecutive single-photon detections using a standard HBT interferometer. After normalization to a Poissonnian statistic following the procedure described in Ref.~\cite{Beveratos_EPJD2002}, the recorded histogram is equivalent to a measurement of the second-order autocorrelation function $g^{(2)}(\tau)$ defined by 
\begin{equation}
g^{(2)}(\tau)=\frac{\langle I(t)I(t+\tau)\rangle}{\langle I(t)\rangle^{2}} \ ,
\label{defg2}
\end{equation}
\begin{figure}[t]
\centerline{\resizebox{0.7\columnwidth}{!}{
\includegraphics{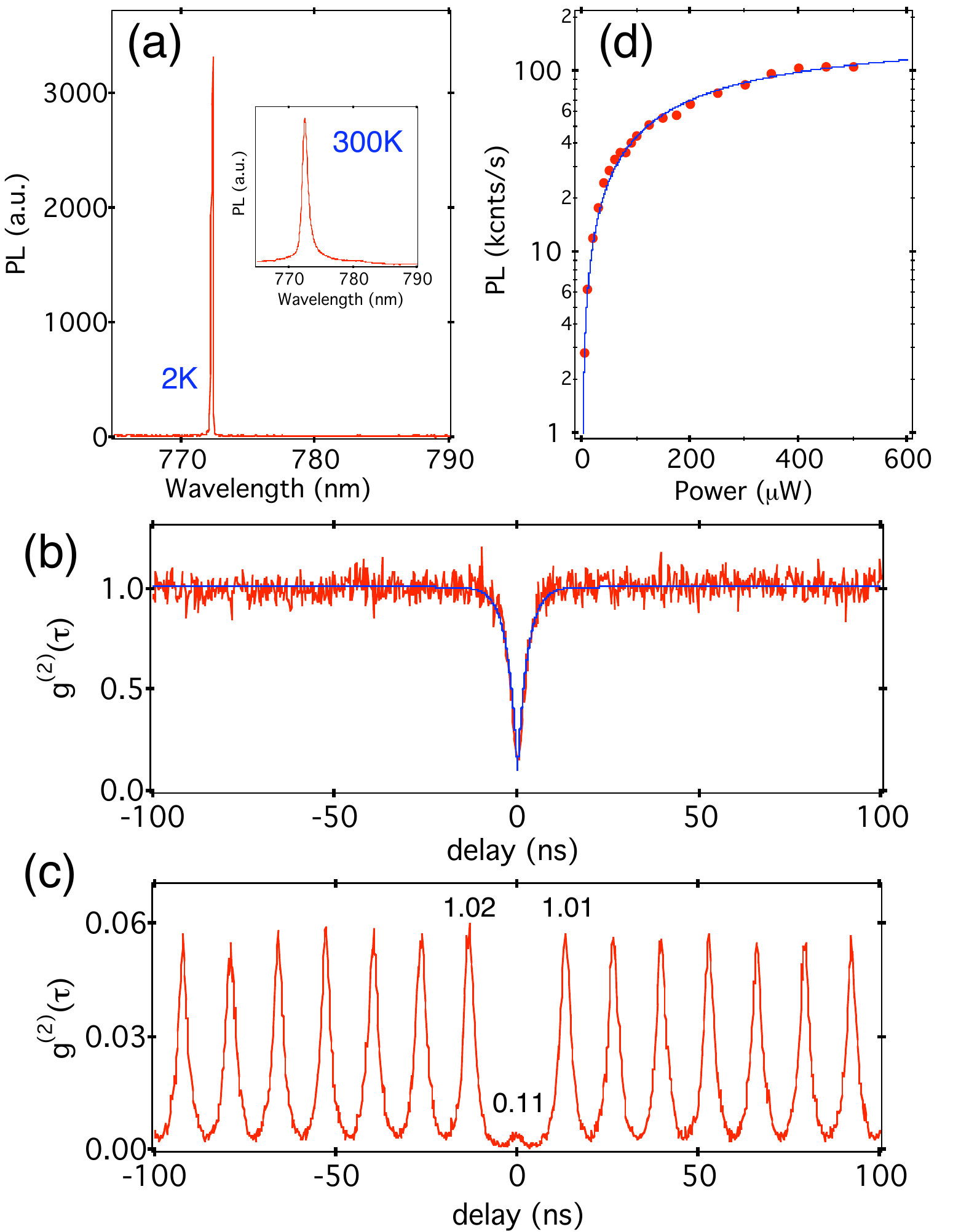}}}
\caption{(a)-PL spectrum of a single defect showing a sharp ZPL in the NIR at the wavelength $\lambda=772$ nm, without any visible phonon sidebands at liquid helium temperature. The width of the ZPL is limited by the resolution of the imaging spectrometer, on the order of $\Delta \nu\approx 100$ GHz. The inset shows the PL spectrum recorded at room temperature, where the first phonon side band can be observed around $780$ nm. (b,c)-Second-order autocorrelation function $g^{(2)}(\tau)$ measured for the same single defect excited at the wavelength $\lambda=700$ nm (b) in a continuous mode and (c) in a femtosecond pulsed regime at a repetition rate of $76$ MHz. Values written above peaks correspond to their respective area after normalization to a pulsed poissonnian light source. (d)- Background-corrected PL intensity as a function of the excitation power. The solid line is data fitting using equation~(\ref{sat}). Data were taken by exciting the defect in a continuous excitation mode. The signal to background ratio is in the order of ten for an excitation power of $300 \ \mu$W.}
\label{Fig2}
\end{figure}
where $I(t)$ is the PL intensity at time $t$. Figure~\ref{Fig2}(b) shows the $g^{(2)}(\tau)$ function recorded while exciting the studied defect in a continuous mode at the wavelength $\lambda=700$ nm. A pronounced anticorrelation effect $g^{(2)}(0)\approx 0.16$ is evidenced at zero delay, which is the signature that a single emitter is addressed. The deviation from a perfect single-photon regime ($g^{(2)}(0)=0$) is due to a residual background PL of the diamond sample, which produces uncorrelated photons associated with Poissonian statistics, and to the detection setup time response function, which is limited by the single-photon detector jitter~\cite{Wu_OptExp2006}.\\
\indent The PL lifetime was then measured by recording the $g^{(2)}(\tau)$ function using femtosecond pulsed excitation at a repetition rate of $76$ MHz. As shown in Fig.~\ref{Fig2}(c), the autocorrelation function exhibits peaks of identical height separated by the repetition period, while the peak at zero delay is missing. After normalization to a pulsed poissonnian light source~\cite{Beveratos_EPJD2002}, the area of the peak at zero delay is found around $g^{(2)}(0)=0.11$. In such a pulsed measurement, this value gives the intrinsic quality of the single-photon source, which is only limited by the signal to background ratio. Fitting each peak of the autocorrelation function with exponential decay, a radiative lifetime $T_{1}=1.96\pm 0.02$ ns was deducted. This value, which is in a good agreement with previously reported lifetime associated with NIR emissions in diamond~\cite{Wu_OptExp2006,Igor_PRB2009}, is almost one order of magnitude shorter than the one associated with NV defect in diamond nanocrystals~\cite{Beveratos_PRA2001}.\\
\indent In order to evaluate the emission rate at saturation $R_{\infty}$, the PL rate $R$ was measured as a function of the laser power $P$ (see figure~\ref{Fig2}(d)). Experimental data were then fitted using the relation
\begin{equation}
R=R_{\infty}\frac{P}{P+P_{\rm sat}} \ ,
\label{sat}
\end{equation}
where $P_{\rm sat}$ is the saturation power, yielding to $R_{\infty}=170$ kcounts.s$^{-1}$. We note that saturation emission rates up to $400$ kcounts.s$^{-1}$ have been measured for single defects emitting in the NIR. Such a variation of the saturation rate among different single emitters is most probably related to different dipole orientations of the defects inside the nanocrystals, corresponding to different efficiencies of light collection. Once again, this results can be compared to the ones obtained for NV defects in diamond. For such emitters, the saturation emission rate is on the order of $50$ kcounts.s$^{-1}$ using the same low temperature experimental setup. Owing to the NV defect Debye-Waller factor ($\approx 0.05$), the counting rate in the ZPL is then around $2$ kcounts.s$^{-1}$, two orders of magnitude smaller than the one associated with the studied defects emitting in the NIR. Such high counting rates indicate that the measured short radiative lifetime is presumably associated to a strong radiative oscillator strength, and not to a PL quenching through fast non-radiative decay processes.\\
\begin{figure}[t]
\centerline{\resizebox{0.7\columnwidth}{!}{
\includegraphics{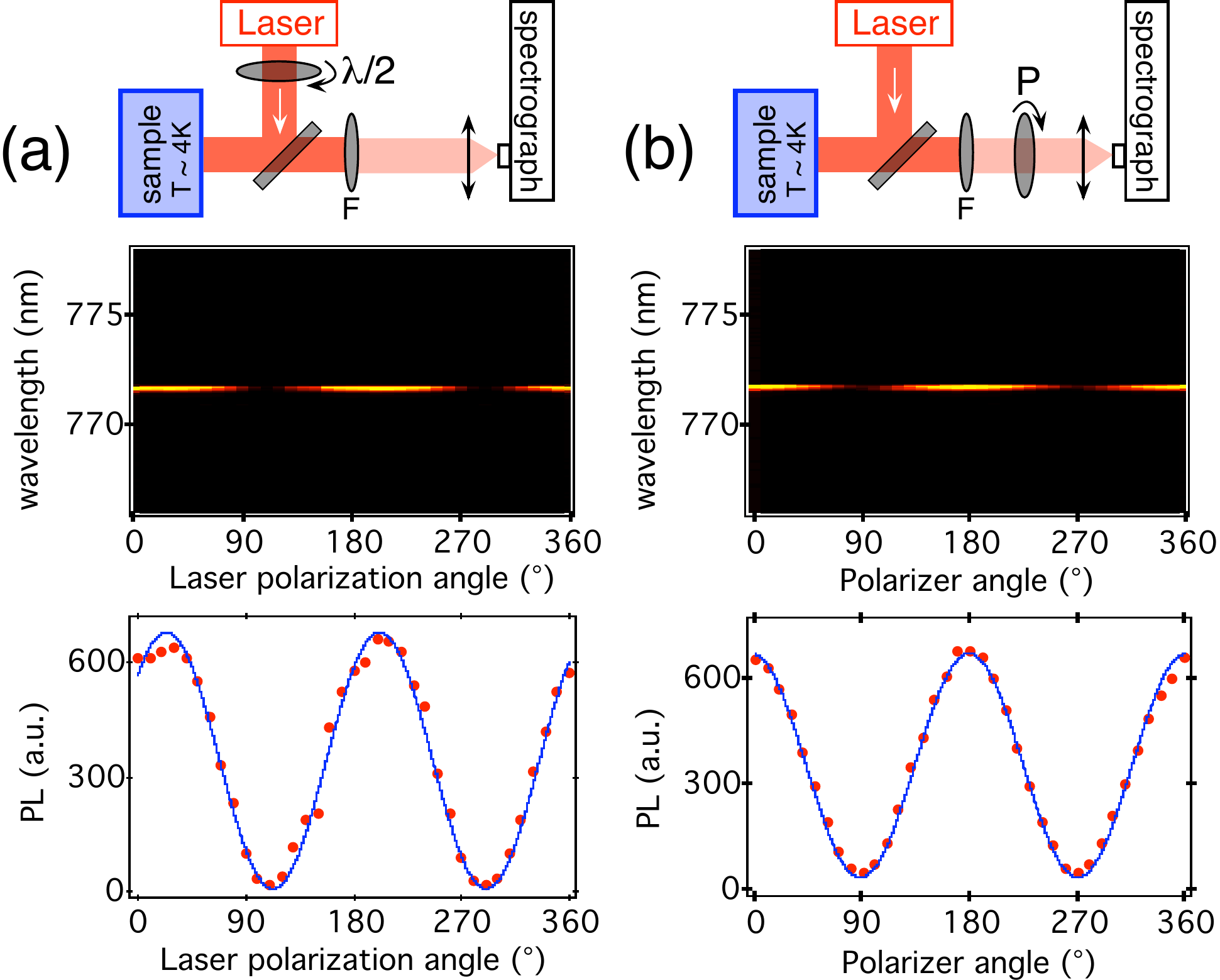}}}
\caption{Polarization properties of a single defect emitting in the NIR in a CVD-grown diamond nanocrystal. (a)-PL intensity as a function of the excitation laser polarization angle, recorded by accumulating emission spectra while rotating the excitation laser polarization with a half-wave plate. The acquisition time for each spectrum is $1$ s. Solid line is data fitting using a Malus-type law, yielding to a contrast of $97\%$. (b)-PL intensity as a function of the angle of a polarizer (P) installed in the detection channel in front of the spectrometer. Solid line is data fitting using a Malus-type law, leading to a contrast of $92\%$. We note that the decrease of the modulation contrast in emission is due to a slight elliptical polarization accumulated along the optical path from the sample to the single-photon detector~\cite{Wu_JLum2006}.}
\label{Fig3}
\end{figure}
\indent We then investigate the polarization properties of the same defect. Figure~\ref{Fig3}(a) depicts an accumulation of PL spectra recorded while rotating the excitation laser polarization. By integrating such spectra, the PL intensity can be displayed as a function of the laser polarization angle. As shown in figure~\ref{Fig3}(a), a modulation with a contrast of $97\%$ was observed, which is the signature that the defect behaves as a perfect single dipole relative to absorption of light. We then studied the polarization properties of the emitted photons by fixing the excitation laser polarization angle parallel to the dipole orientation, and by introducing a polarizer in front of the spectrometer. Following the method described above, PL spectra were then accumulated while rotating the polarizer in the detection channel (see figure~\ref{Fig3}(b)). Once again, a modulation with a contrast close to unity was observed indicating that the defect also behaves as a single emitting dipole, in opposition to NV defects where two orthogonal dipoles are always involved~\cite{Epstein_NatPhys2005,Kaiser_QuantPh2009}. We note that the polarization properties are very important for the use of single-photon emitters in practical quantum key distribution applications, where the information can be encoded in the polarization of the single photons~\cite{GisinRevue}. \\

\begin{figure}[t]
\centerline{\resizebox{0.7\columnwidth}{!}{
\includegraphics{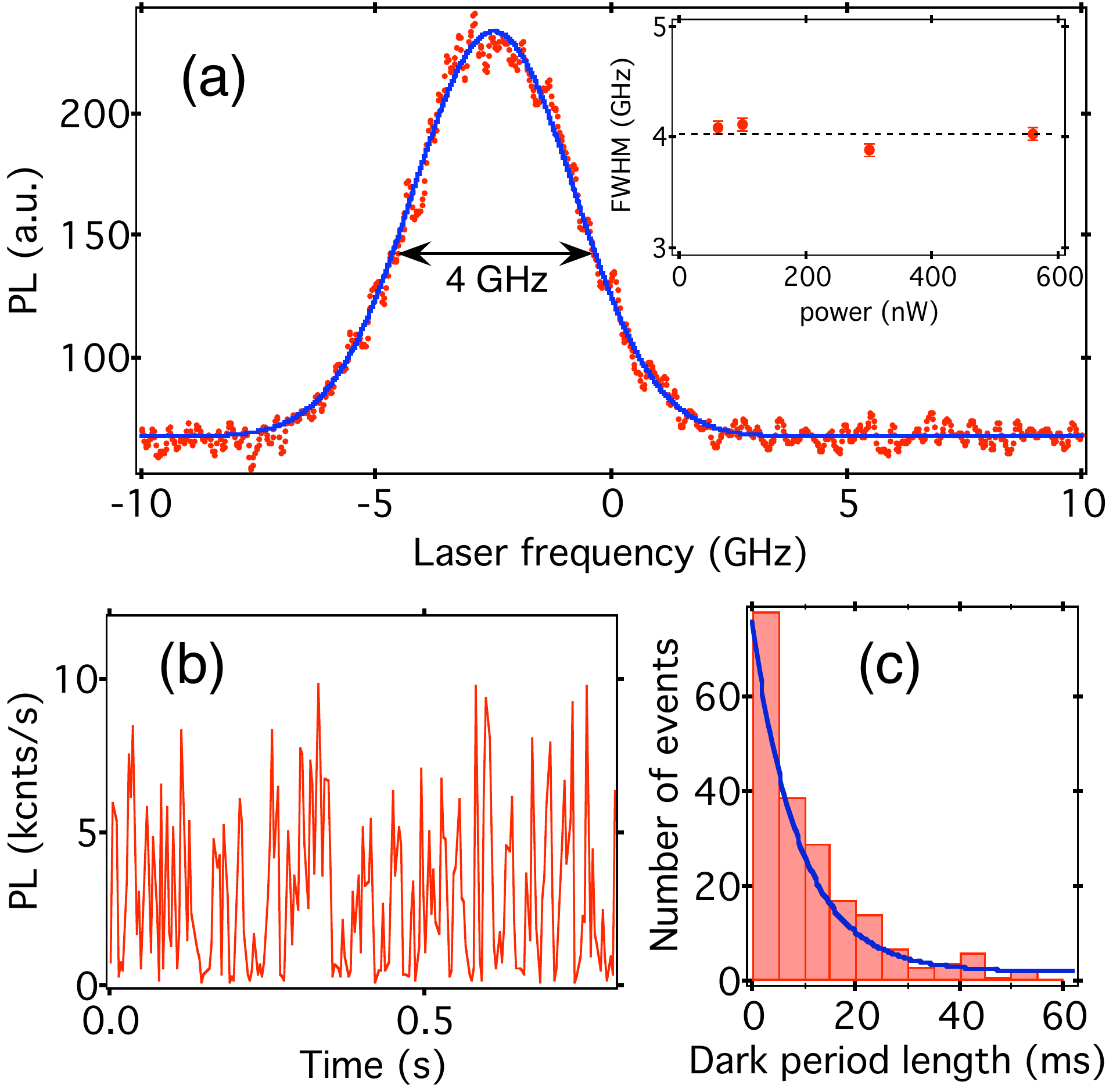}}}
\caption{(a)-Resonant excitation spectrum of a single defect emitting in the NIR in a CVD-grown diamond nanocrystal. For this emitter, the ZPL is centered at $760$~nm. The total integration time is $16$~mn, corresponding to 120 sweeps of the laser frequency. The solid line is data fitting using a Gaussian profile and the inset shows the evolution of the optical resonance linewidth as a function of the excitation power. (b)-PL time-trace recorded while exciting the defect at resonance with a power of $400$~nW. The integration time per point is $5$~ms. The observed blinking behavior is the signature of slow spectral jumps. (c)-Histogram of the length of the dark periods in the PL-time trace over an observation time of $5$~s. The solid line is data fitting using a single exponantial decay, giving a spectral diffusion time of $10\pm 2$ms.}
\label{Fig4}
\end{figure}

\indent As already mentioned in the introduction, most of the applications of single-photon source in the field of quantum information processing requires the emission of indistinguishable single photons~\cite{RevueOrrit,KLM,OBrien_Science2007}, {\it i.e.} photons in the same spatial mode, with identical polarization and with a Fourier-transform limited relation between their spectral and temporal profiles. So far, generation of indistinguishable single photons have been achieved using parametric down conversion~\cite{HOM}, single trapped ions~\cite{Monroe_NatPhys2007,Eschner_NJP2009} or atoms~\cite{Rempe_PRL2004,Grangier_Nature2006}, single molecules~\cite{Zumbusch_PRL2005,Vahid_OptExp2007}, single semiconductor quantum dots~\cite{Santori_Nature2002,Abram_APL2005,Bennett_APL2008} and NV defects in diamond~\cite{Tamarat_PRL2006,Batalov_PRL2008}. The controlled generation of indistinguishable single photons in solid-state systems remains however challenging because strong interactions with the host matrix partially destroy the coherence between consecutive emitted single photons. In the following, we investigate the spectral stability of single defects emitting in the NIR in diamond nanocrystals, in order to check if Fourier-transform limited emission can be achieved. For that purpose, a frequency-stabilized single mode tunable laser (Ti:Sa) with a linewidth smaller than $1$ MHz and a mode-hop free frequency tuning around $20$~GHz was used to excite resonantly the defects on their ZPL. The red-shifted PL was detected using a $785$ nm long-pass filter and monitored while sweeping the single-mode laser frequency. As the Debye-Waller factor is high, the rate of red-shifted PL 
is low, typically on the order of a few kcounts.s$^{-1}$.\\
\indent A resonant excitation spectrum of a single defect is depicted in figure~\ref{Fig4}(a). The data are well fitted by a Gaussian profile with a full-width at half maximum (FWHM) around $4$ GHz. Owing to the previously measured radiative lifetime $T_{1}=2$ ns, the lifetime-limited linewidth $\Delta \nu_{\rm FT}$ is given by~\cite{{RevueOrrit}}
\begin{equation}
\Delta \nu_{\rm FT}=\frac{1}{2\pi T_{1}} \approx 80 \ {\rm MHz} \ .
\label{LT}
\end{equation}
Therefore, the excitation linewidth is not Fourier transform limited. We note that the broadening of the optical resonance does not arise from power broadening as the emission linewidth does not change while increasing the excitation power, as shown in figure~\ref{Fig4}(a).\\
\indent Spectral broadening results from fluctuations of the optical resonance frequency, either by dephasing or by spectral diffusion~\cite{RevueOrrit}. Pure dephasing processes are fast, and arise from interactions with phonons in the crystalline matrix. Such processes lead to an homogeneous broadening which follows a Lorentzian profile. Owing to the measured Gaussian profile of the optical resonance, we tentatively conclude that dephasing is not the predominant broadening process. Spectral diffusion leads to comparatively stronger fluctuations, associated with slow frequency drifts and jumps, which can be evidenced by recording the red-shifted PL time-trace while exciting the defect at resonance. As shown in figure~\ref{Fig4}(b), a strong blinking behavior is observed which is the signature of slow spectral jumps. The histogram of the length of the dark periods is shown in  figure~\ref{Fig4}(c). Data fitting with a single exponantial decay gives an estimate of the spectral diffusion time $T_{\rm sd}=10\pm 2$ ms, much longer than the radiative lifetime $T_{1}$. We note that if two consecutive photons are emitted within a time interval shorter than the characteristic spectral diffusion time, these two photons might be indistinguishable, as the slow spectral diffusion process could be neglected~\cite{Santori_Nature2002}. Consequently, owing to a spectral diffusion time in the millisecond range, it should be possible to use defects emitting in the NIR in diamond nanocrystals to perform a two photon interference experiment, which is at the heart of linear quantum computing protocols~\cite{KLM}. Although not observed to date, this preliminary data would indicate that further investigation is warranted.\\
\indent Since the atomistic structure of the defect is still unknown, it is hard to identify the physical processes leading to spectral jumps. One may consider a charge hoping the defect on/off, therefore changing its emission line, or a fluctuating strain at the location of the defect, which is known to change drastically the energy level structure of color centers in diamond~\cite{Batalov_PRL2009}. Furthermore, previous studies on NV defects in diamond have shown that strain effects are much stronger in diamond nanocrystals than in bulk samples~\cite{Jelezko_JMol2001,Wang_PRB2008}. Indeed, NV defects usually exhibit poor spectral stability in diamond nanocrystals while Fourier-transform emission can be achieved in bulk samples~\cite{Tamarat_PRL2006,Batalov_PRL2008}. Therefore, the spectral stability might be greatly improved by working with bulk diamond samples.\\
\begin{figure}[b]
\centerline{\resizebox{0.75\columnwidth}{!}{
\includegraphics{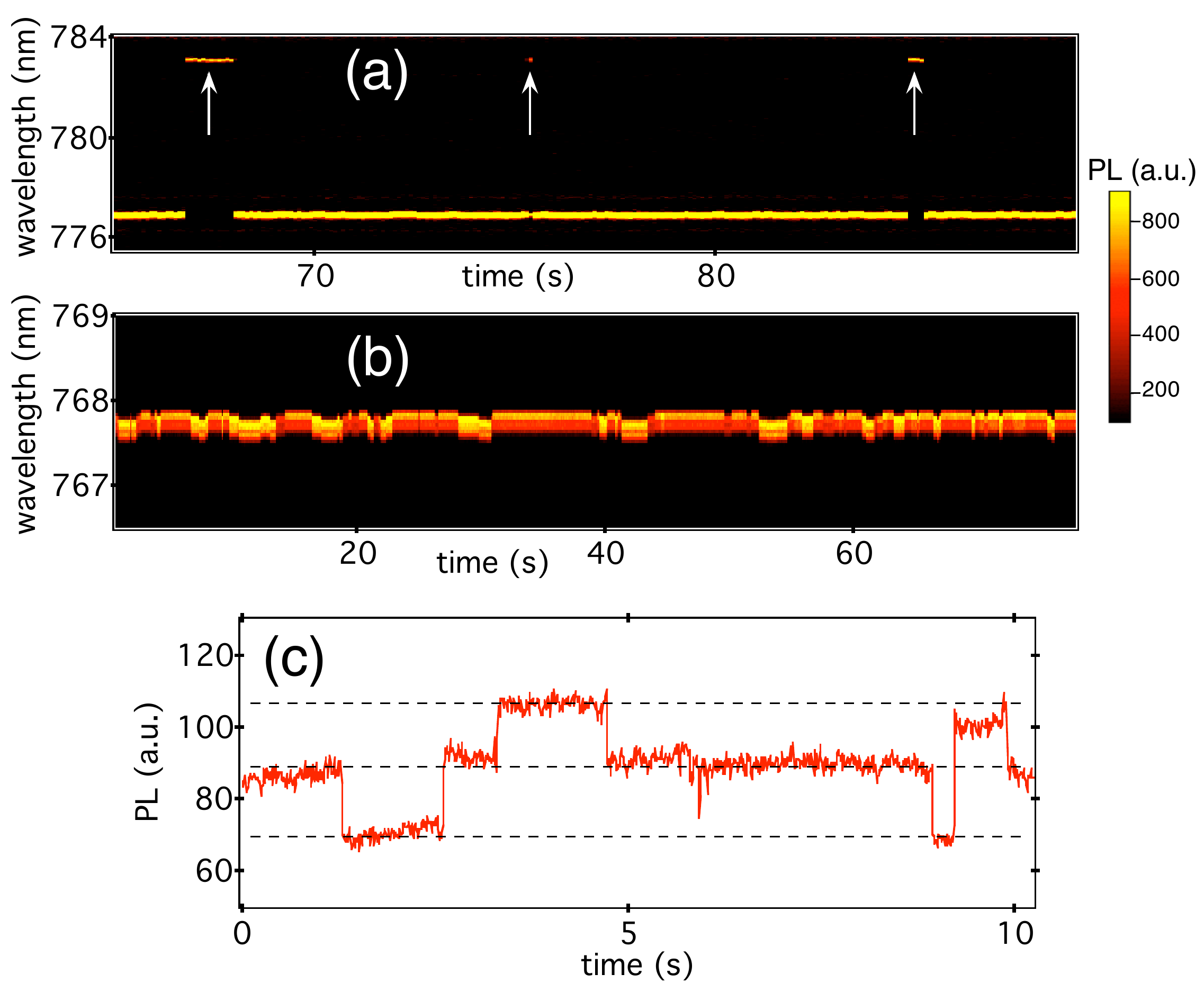}}}
\caption{(a)-Real-time acquisition of PL spectra recorded for another type of single  defect. Spectral jumps up to $6$ nm are observed (white arrows). Each spectrum is accumulated during $100$ ms. (b),(c)-PL spectra and time-trace recorded from another  defect emitting in the NIR. A blinking behavior is correlated to spectral jumps. For this particular defect, three different emission states can be observed from the different counting rates of the center (dash lines).}
\label{Fig5}
\end{figure}
\indent We have investigated around $25$ diamond nanocrystals hosting single defects with a ZPL around $770$ nm, within a range of $770\pm15$ nm. We note that permanent photobleaching has been observed for two defects while exciting in a pulsed regime. Furthermore, we stress that only $20\%$ of the studied defects were exhibiting an optical resonance linewidth in the GHz range. For most of the studied defects, high amplitude spectral jumps ($>100$ GHz) could be observed in real-time acquisition of PL spectra, as depicted in figure~\ref{Fig5}(a) and (b). For this class of single emitters, a blinking correlated to spectral jumps was also evidenced in the PL time-trace. As an example, figure~\ref{Fig5}(c) shows a correlate PL time-trace for the spectrum shown in figure~\ref{Fig5}(b) . Such spectral jumps might be associated with a modification of the defect configuration. Using a single mode laser with a mode-hop free frequency tuning around 20 GHz, it was not possible to record resonant excitation spectra for this type of emitters. \\

\indent To summarize, we have reported a study of the optical properties of single defects emitting in the NIR in CVD-grown nanodiamonds at liquid helium temperature. Such defects exhibit several striking features: (i) a sharp PL in the NIR associated with a high Debye-Waller factor ($>0.9$), (ii) high counting rates, (iii) a short radiative lifetime ($2$~ns), and (iv) a perfectly linearly polarized emission. In addition, an optical resonance linewidth of $4$ GHz is reported using resonant excitation on the zero-phonon line. The broadening of the optical resonance results from spectral jumps ocurring at the millisecond timescale. Such results are promising for the future realization of an efficient source of indistinguishable single photons using single defects in diamond. However, it will be highly desirable in future to control the creation of such defects in bulk samples in order to improve their spectral stability and to investigate in details their atomistic structure. In particular, the spin structure of the defect could be investigated by recording resonant excitation spectra while applying a magnetic field. Owing to the coherence properties of spin states in diamond~\cite{Gupi_NatMat2009}, such experiments could open many perspectives in the context of coupling between spin states and optical transitions.

\ack{We acknowledge financial support by BMBF (projects KEPHOSI and EPHQUAM), the European Union (QAP, EQUIND (Project No. IST-034368), and NEDQIT), the Landesstiftung Baden-W$\ddot{\rm u}$rttemberg, the European Research Council (FP7/2007-2013)/ERC Grant agreement No. 209636, the Australian Research Council and The International Science Linkages Program of the Australian Department of Innovation, Industry, Science and Research. V J is supported by the Humboldt Foundation.}

\end{document}